# Effective tuning methods for few-electron regime in gate-defined quantum dots

Chanuk Yang[1§], Hwanchul Jung[2§], Hyung Kook Choi[1,3*], Yunchul Chung[2†]

[1] Department of Physics, Research Institute of Physics and Chemistry, Jeonbuk National University, Jeonju 54896, Republic of Korea

[2] Department of Physics, Pusan National University, Busan 46241, Republic of Korea

[3] K-tip Corporation, Jeonju 54896, Republic of Korea

§: These authors contributed equally to this work.

We present systematic methods for compensating gate crosstalk effects in gate-defined quantum dots (QDs), to allow the observation of Coulomb blockade peaks from the few-electron regime ($N = 1$) to $N \approx 20$. Gate crosstalk, where adjustments to one gate voltage unintentionally affect other gate-controlled parameters, makes it difficult to control tunneling rates and energy states of the QD separately. To overcome this crosstalk effect, we present two approaches: maintaining constant conductance of two quantum point contacts (QPCs) forming the QD by compensating the effect of the plunger gate voltage on the QPCs, and interpolating between gate voltage conditions optimized for QD observation at several electron numbers. These approaches minimize crosstalk effects by dynamically adjusting barrier gate voltages as a function of plunger gate voltage. Using these methods, we successfully observed Coulomb blockade peaks throughout the entire range from $N = 1$ to $N \approx 20$. Our methods provide a simple and effective solution for observing Coulomb blockade peaks over a wide range of electron numbers while maintaining control over the quantum states in the dot.

Keywords: Quantum dot, crosstalk, gate compensation

Email: * **hkchoi@jbnu.ac.kr** and † **ycchung@pusan.ac.kr**

Fax: +82-63-270-3320

# Introduction

Gate-defined quantum dots (QDs) in two-dimensional electron systems have proven to be a versatile and powerful platform for exploring diverse quantum phenomena, including the Coulomb blockade [1], Kondo effect [2–5], charge frustration [6], spin-orbit interactions [7], quantum interference [8–10], valley physics [11], exchange interactions [12], and Pauli spin blockade [13] etc. These phenomena offer insights into the fundamental interactions governing nanoscale systems and the emergence of quantum correlations. QD is particularly valued for its precise tunability and scalability, by controlling multiple key parameters: the barrier gates control tunnel coupling with the electron reservoirs; the plunger gate adjusts the energy levels within the QD, which allows for systematic manipulation of the QD's electronic states and its interaction with the surrounding environment.

Recently, QDs in multi-dot configurations have shown great potential as quantum simulators [14,15] and spin qubits [16,17], utilizing the spin of a confined electron or hole as a quantum bit. Furthermore, research on flying qubits has demonstrated that quantum information can be coherently transferred between spatially separated QDs [18]. These capabilities establish QDs as crucial building blocks for applications in quantum technologies.

In many research efforts, the ability to reliably tune QDs in the few-electron regime ($N = 1\sim20$) is crucial. However, tuning the device to observe well developed Coulomb blockade peaks in the few-electron regime remains challenging due to the inherent crosstalk between different gates that control the quantum dot [19]. Traditional tuning approaches often struggle to maintain consistent tunnel coupling of the dot across wide ranges of electron numbers, limiting our ability to perform experiments that require precise control of electron numbers and tunnel couplings. The challenge arises from the complex interplay between gate tuning parameters. Specifically, adjusting the plunger gate voltage to control the dot's electron number influences the tunnel barriers inevitably, even when barrier gate voltages are fixed. Such crosstalk often leads to unintended outcomes, such as complete pinch-off of the QD or excessive coupling to the reservoirs, thereby preventing the clear and consistent observation of Coulomb blockade peaks across the entire few-electron regime. This limitation severely hampers the study of quantum phenomena, which require precise control of electron numbers and tunnel barriers. In this study, we propose two novel approaches to compensate for gate crosstalk effects. These approaches enable observation of Coulomb blockade peaks over a wide range of electron numbers from $N = 1$ to $N \approx 20$ while maintaining control over the quantum states in the dot.

## Experimental Setup and Results

The experiments were conducted using devices fabricated on a GaAs/AlGaAs heterojunction, with a two-dimensional electron gas (2DEG) formed 70 nm below the wafer surface. Metallic Schottky gates were deposited to define both the QD and its associated charge sensor, as shown in Figure 1(a). The QD is controlled by four gates. The Left ($V_L$) and Right ($V_R$) barrier gates control coupling with the reservoirs, while the Plunger gate ($V_P$) adjusts the QD energy levels. The Up gate ($V_U$) tunes the capacitive coupling between the QD and charge sensor. The charge sensor is defined by an upper quantum point contact gate ($V_{QCS}$). Transport measurements were performed using standard lock-in techniques with custom-built transimpedance amplifiers. For QD conductance measurements, we applied an AC excitation voltage of 10 μV at 456.78 Hz, while the charge sensor was measured at 345 Hz to avoid crosstalk between the two adjacent measurement channels.

Figure 1(b) presents measurement results using a conventional approach. The blue trace shows Coulomb blockade peaks obtained by sweeping the plunger gate voltage while keeping the left and right QPC gate voltages fixed. The red trace represents measurements from the charge sensor, which detects changes in the electron occupation of the QD. The charge sensor data reveals that the last electron tunnels out of the QD at the plunger gate voltage of approximately -0.73 V. As we sweep the plunger gate voltage, each sharp change in the charge sensor conductance indicates a single-electron tunneling event, enabling us to track the electron number in the QD throughout the entire voltage range. No Coulomb blockade peaks are observed when the electron number is less than four. This is due to the extremely weak tunneling between the QD and the 2DEG reservoirs, resulting in current through the QD being too small to measure. Coulomb blockade peaks become observable for electron numbers $N \geq 4$. In particular, the peaks corresponding to $N = 4, 5$, and 6 show clear separation with minimal background conductance between adjacent peaks. As the plunger gate voltage is increased towards more positive values, progressive broadening of the Coulomb blockade peaks is observed. These peaks gradually overlap with each other, leading to substantial background conductance increase and ultimately resulting in background conductance that exceeds the peak heights. This is because crosstalk causes enhanced tunneling through the QPCs that define the QD as the plunger gate voltage is increased towards more positive values. Consequently, the broadening of energy levels in the QD becomes larger than the energy level spacing, resulting in overlapping Coulomb blockade peaks and increased background conductance.

Figure 1(c) schematically illustrates how crosstalk from the plunger gate modifies the QD potential profile. Moving from right to left in the figure, the QPC barriers that define the QD become more opaque due to crosstalk as the plunger gate voltage is made more negative, even with fixed QPC barrier gate voltages. Given that the separation between gates and the two-dimensional electron gas is approximately 70 nm, while the distance between gates is approximately 150 nm, the electric field from the plunger gate inevitably affects the electron gas underneath the QPC barrier gates too. Specifically, making the plunger gate voltage, $V_P$ more negative to reduce electron numbers simultaneously affects the QPC tunnel barriers due to the fringing fields from the plunger gate, even when QPC barrier gate voltages are fixed. This crosstalk hampers the observation of Coulomb blockade peaks over a wide range of electron numbers using conventional fixed-barrier voltage approaches. Since this effect is intrinsic to devices with relatively thick insulating layers between gates and the conducting channel, we developed compensation techniques as described below.

To acquire the crosstalk compensation data, we measured the influence of the plunger gate voltage on the left and right QPCs separately. For the measurement, the left (right) QPC conductance was measured as a function of the left (right) QPC voltage and the plunger gate voltage, while setting the right (left) QPC fully open. The influence of the plunger gate voltage on the left and right QPCs is presented in Figures 2(a) and 2(b), respectively. Both measurements reveal strong dependence of QPC conductance on the plunger gate voltage. To compensate for the crosstalk, we determined the required QPC gate voltages that maintain a fixed barrier transmission of 0.3 $G_0$ ($G_0 = e^2/h$ is the conductance quantum, where $e$ is electron charge and $h$ is Planck constant) independent of plunger gate voltage (allowing for an error margin of 0.01 $G_0$). These voltages were determined to counteract the changes in plunger gate voltage.

In these two-dimensional data sets, the points corresponding to conductance of 0.3 $G_0$ are marked as black open circles in Figure 2(a, 65 data points) and Figure 2(b, 32 data points). The obtained data points are discrete with respect to the plunger gate voltage. To implement continuous compensation during experiments, where the plunger gate voltage is swept continuously, we need to derive a continuous function that determines the required QPC gate voltages to maintain the conductance at 0.3 $G_0$ throughout the entire plunger gate voltage range. Three different methods were applied to derive the continuous function: (i) Modified Akima cubic Hermite interpolation (Makima), which provides a smoother curve with reduced overshooting, represented by green dots; (ii) second-order polynomial fitting, indicated by a blue solid line; and (iii) third-order polynomial fitting, shown by a red dashed line.

The Coulomb blockade peaks measured after crosstalk compensation show clear separation with vanishing background conductance, as demonstrated in Figure 2(c-e) using compensation derived by Makima interpolation, second-order polynomial fit, and third-order polynomial fit, respectively. Furthermore, compared to the uncompensated measurement shown in Figure 2(b), the charge sensing signal shows distinct conductance steps corresponding to single-electron changes in the QD occupation, maintaining high visibility even at high electron numbers. The enhanced measurement clarity can be attributed to the following mechanism: Without compensation, increasing the plunger gate voltage leads to enhanced tunneling rates, causing energy level broadening in the QD that exceeds the level spacing. Consequently, the Coulomb blockade peaks overlap and electron occupation cannot be precisely determined at specific plunger gate voltages, resulting in reduced charge sensor visibility. In contrast, maintaining weak tunnel coupling through compensation preserves well-defined sharp energy levels in the QD, enabling precise determination of electron occupation throughout the plunger gate voltage range which enhances the charge sensor visibility.

While all three fitting methods yield satisfactory results, they exhibit a common limitation: the Coulomb blockade peak corresponding to $N = 1$ is either significantly suppressed or barely detectable (vertical dot guideline in Figure 2(c, d, e)). This limitation stems from two effects in the $V_P < -0.8$ V regime. First, as the QPCs approach their pinch-off region, the barrier transmission becomes extremely sensitive to small variations in the plunger gate voltage. Second, unwanted resonances appear, making it challenging to identify points with conductance of 0.3 $G_0$. These resonances arise from our device cooling procedure, where positive gate voltages were applied to enhance stability. As a result, residual potential barriers persist at the opposite QPC even at high positive gate voltages, causing partial reflection of electrons transmitted through the QPC under measurement. This reflection creates resonances that prevent accurate conductance measurements of individual QPCs in this regime.

To overcome this limitation, we carefully adjusted all gate voltages to maximize the visibility of the first electron Coulomb blockade peak, achieving an observable peak with conductance of 0.04 $G_0$. As shown in Figure 3(a, b), these optimized gate voltages (black dot) represent a slight modification from the original values. The previously obtained second and third-order polynomial functions were then adjusted using two different shift methods. The first method involved a horizontal shift ($\Delta V_P$) to align the functions with the plunger gate voltage where the first Coulomb blockade peak is observed. The second method employed a vertical shift ($\Delta V_{L,R}$) to make the functions pass through the optimized left and right gate voltages that produce the first Coulomb blockade peak. Subsequent conductance measurements were performed using both modified second and third-order polynomial fitting curves incorporating these adjustments, with the results presented in Figure 3(c) and (d). As demonstrated in Figure 3(c, d), the horizontal shift of the polynomial functions resulted in well-resolved Coulomb blockade peaks throughout the measurement range. The vertical shift of gate voltages, however, suppresses the Coulomb blockade peaks at higher electron occupancy, which is not optimal for measurements. In the present work, we limited our approach to simple shifts of the fitting curves. An alternative method would be to derive entirely new polynomial functions by replacing the gate voltage values in the pinch-off region with the conditions optimized for the first Coulomb blockade peak.

The second approach presents a different strategy from the previous method of measuring and compensating for the plunger gate's influence on individual QPC conductances. Rather than tracking conductance changes, we first determined the optimal gate voltage conditions that produce well-defined Coulomb blockade peaks at specific electron numbers ($N$ = 1, 5, and 16), then used these optimal values to derive polynomial functions for crosstalk compensation.

To find the conditions for optimal Coulomb blockade peak formation, we systematically varied the left and right gate voltages that define the quantum dot and plotted the Coulomb blockade peak heights at three different electron numbers $N$ = 1, 5, and 16 in two-dimensional plots, as shown in Figure 4(a-c). The color of each pixel represents the maximum conductance of the quantum dot (QD) at a particular $V_L$ and $V_R$.

The highest conductance peaks appear along the diagonal direction in the $V_L$ vs. $V_R$ plot, where symmetric tunneling rates through the left and right barriers are achieved. Based on both the height and shape of the conductance peaks, three optimal sets of $V_L$, $V_R$ and $V_P$ parameters were selected. These optimal points are marked by open circles in Figures 4(a-c) (black for $N$ = 1, blue for $N$ = 5, and red for $N$ = 16) and plotted in $V_{L,R}$ vs. $V_P$ space in Figure 4(d, e). Using these three selected points, second-order polynomial fittings were performed on the conductance plots of $V_L$ vs. $V_P$ and $V_R$ vs. $V_P$, as indicated by the blue lines in Figure 4(d, e). Figure 4(f) displays the QD conductance measured along this fitting curve, showing distinct Coulomb peaks.

Figure 4(f) shows the Coulomb blockade peaks measured using the above-mentioned compensation method. Using this compensation method, we obtained similar results to those shown in Figures 2 and 3. The polynomial fitting functions derived in Figure 4(d, e) exhibit only minor differences from the previously obtained functions. However, these small variations produce distinctly different Coulomb blockade peak patterns, highlighting the extreme sensitivity of peak formation to precise gate voltage conditions. The primary advantage of this approach is that it enables us to optimize the Coulomb blockade peak shapes and quantum dot states for specific electron numbers of interest. The trade-off, however, is the considerable time required to generate the two-dimensional conductance maps by scanning the QPC barrier gate voltages.

## Conclusion

In this study, we have demonstrated two systematic approaches to compensate the gate crosstalk in gate-defined quantum dots, particularly in the few-electron regime. The first method compensates for crosstalk by maintaining constant QPC conductance as the plunger gate voltage varies, achieved through polynomial fitting of equi-conductance points. The second method derives compensation functions from gate voltage conditions optimized for specific electron numbers where well-defined Coulomb blockade peaks are observed.

Both approaches successfully enable the observation of Coulomb blockade peaks from $N = 1$ to $N \approx 20$, while maintaining well-controlled tunnel coupling. The first method provides comprehensive coverage across the entire measurement range but requires careful handling in the few-electron regime where QPCs approach their pinch-off region. The second method allows precise control of quantum dot states at specific electron numbers of interest, though it requires time-consuming two-dimensional conductance mapping.

Our results validate the effectiveness of these compensation techniques in maintaining stable quantum dot operation across a wide range of electron numbers. The observed sensitivity of Coulomb blockade peaks to small variations in the compensation functions emphasizes the importance of precise gate voltage control. These methods provide practical solutions for quantum dot experiments requiring precise control of both electron numbers and quantum states, particularly beneficial for applications in quantum computation and simulation where reliable few-electron state control is essential.

Future work could focus on automating these compensation procedures to enhance their efficiency. Additionally, extending these techniques to multi-dot systems could facilitate the development of more complex quantum circuits and scalable quantum architectures.

## Acknowledgement

## DATA AVAILABILITY

The data that support the findings of this study are available from the corresponding author upon reasonable request.

**Figure**

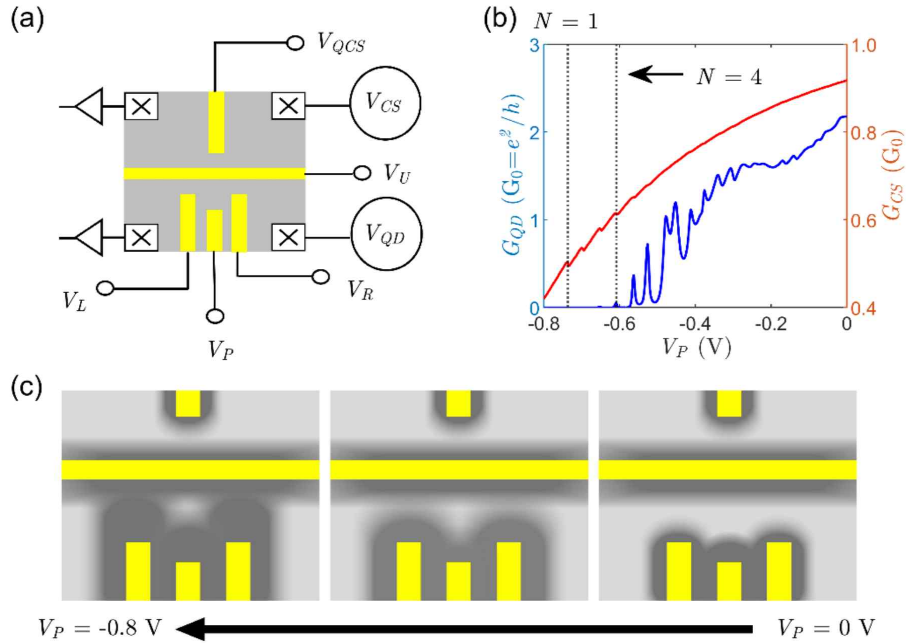

**Figure 1: Device schematic and transport characteristics** (a) Schematic illustration of the gate-defined quantum dot device. The quantum dot is defined by Left ($V_L$), Right ($V_R$), and Plunger ($V_P$) gates, while the charge sensor is controlled by $V_{QCS}$. (b) Transport measurements through the QD (blue) and charge sensor (red) as a function of plunger gate voltage. The blue trace shows Coulomb blockade peaks in conductance measurements, while the red trace exhibits distinct steps corresponding to electron number transitions in the QD. (c) Schematic illustration of how the plunger gate crosstalk affects the QD potential profile. As plunger gate voltage becomes more negative, the QPC barriers become more opaque due to crosstalk, even with fixed barrier gate voltages.

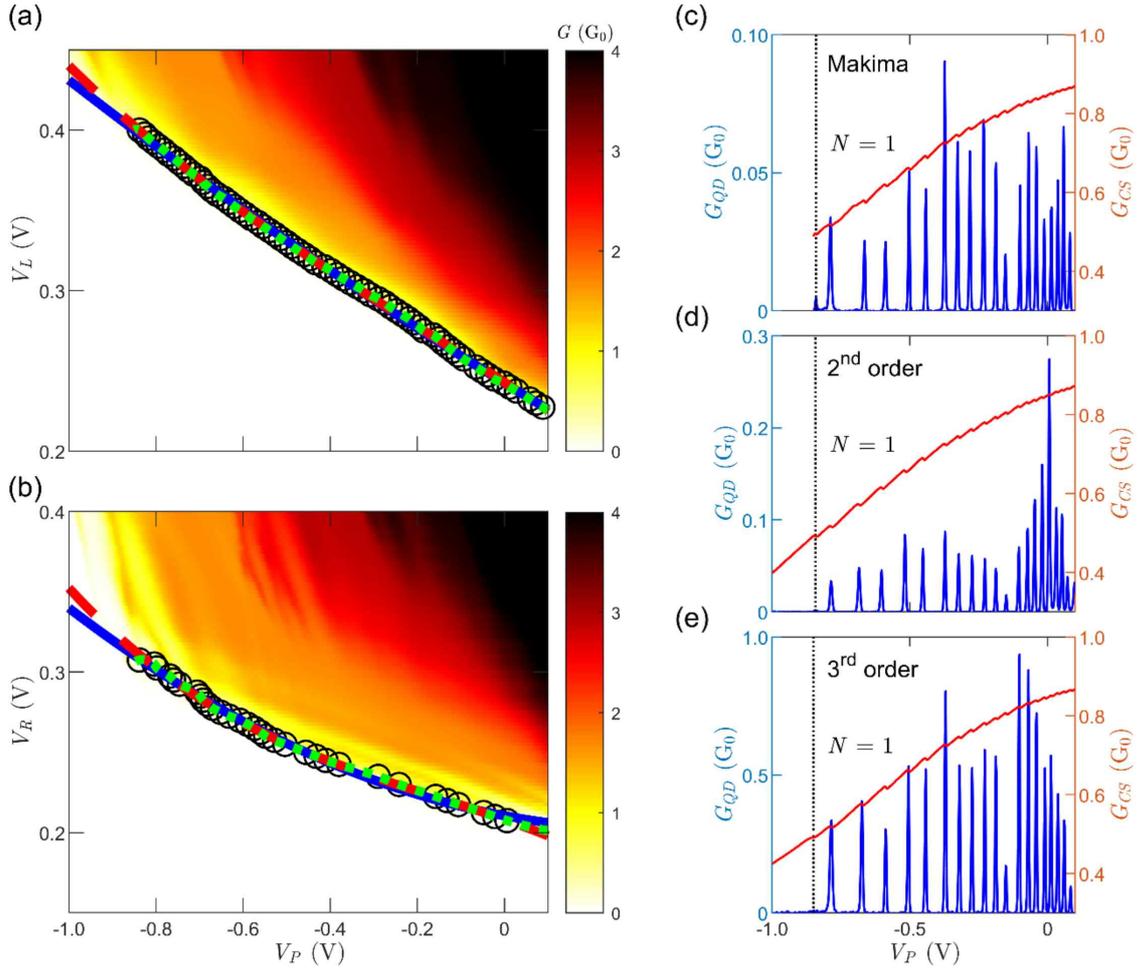

**Figure 2: Equi-conductance tracking method for crosstalk compensation** (a, b) Two-dimensional plots of QPC conductance measured as a function of gate voltage ($V_{L,R}$) and plunger gate voltage ($V_P$) for the left (a) and right (b) QPCs. Black open circles indicate points of constant conductance (0.3 $G_0$) used for deriving compensation functions. (c-e) Coulomb blockade measurements after implementing the crosstalk compensation derived using three different methods: (c) Modified Akima cubic Hermite interpolation (green dots), (d) second-order polynomial fitting (blue solid line), and (e) third-order polynomial fitting (red dashed line). Both QD conductance (blue) and charge sensor response (red) are shown.

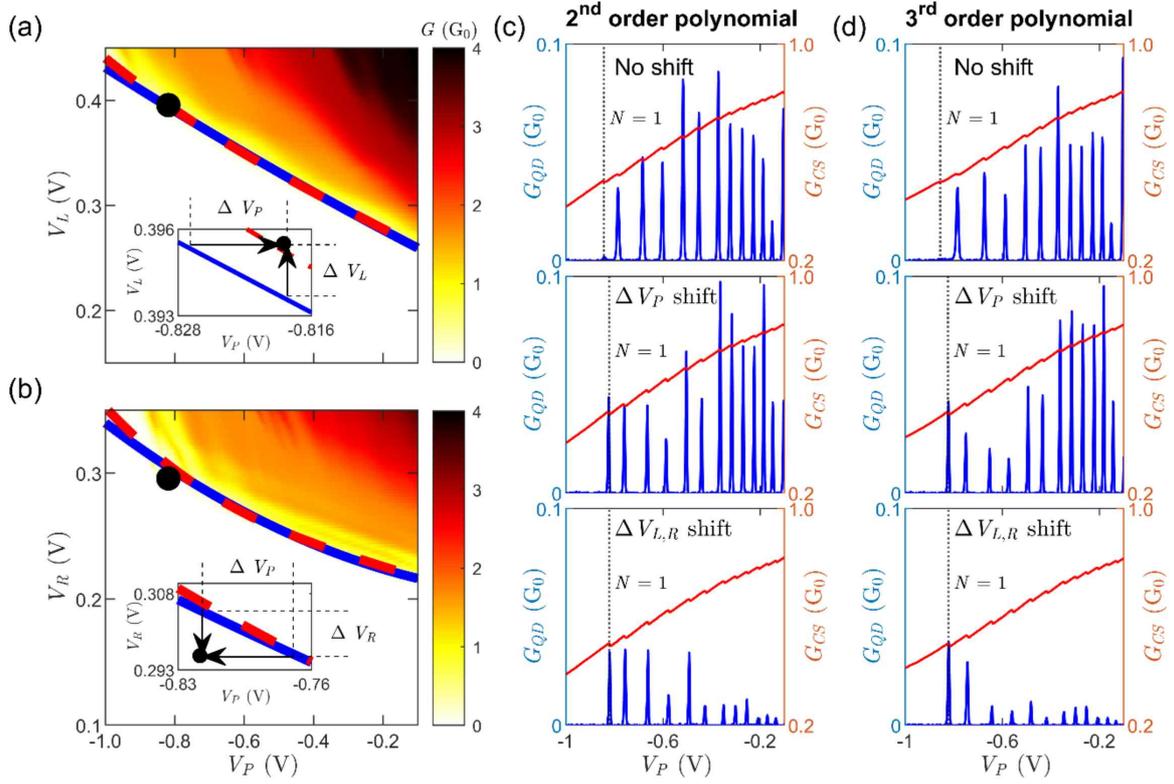

**Figure 3: Modified compensation schemes for single-electron regime** (a, b) Adjustment of compensation functions for left (a) and right (b) QPCs, optimized to observe the first electron state. Black dots represent the modified gate voltage conditions achieving a conductance of 0.04 $G_0$. The inset illustrates two adjustment methods: horizontal shift ($\Delta V_P$) of plunger gate voltage and vertical shift ($\Delta V_{L,R}$) of QPC gate voltages. (c, d) Transport measurements after implementing (c) horizontal and (d) vertical shifts of the compensation functions, showing both QD conductance (blue) and charge sensor response (red).

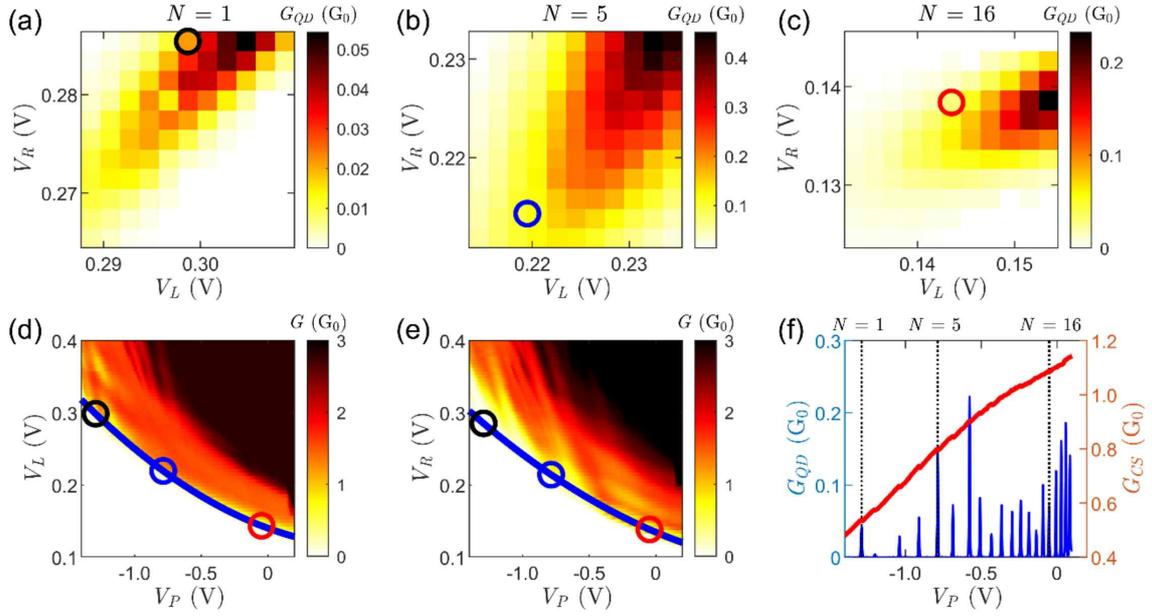

**Figure 4: Crosstalk compensation based on optimal conditions at three electron numbers** (a-c) Two-dimensional conductance maps in $V_L$-$V_R$ space showing Coulomb blockade peak heights for three different electron numbers: (a) $N = 1$, (b) $N = 5$, and (c) $N = 16$. Optimal gate voltage conditions are marked by open circles (black, blue, and red respectively). (d, e) Polynomial compensation functions (blue lines) derived from these three optimal points, showing the required (d) left and (e) right gate voltages versus plunger gate voltage. (f) Transport characteristics measured using the derived compensation functions, showing both QD conductance (blue) and charge sensor response (red).